\documentclass[preprint,showpacs,preprintnumbers,amsmath,amssymb]{revtex4-1}
\usepackage{mathrsfs}
\usepackage{graphicx}
\usepackage{dcolumn}
\usepackage{bm}

\begin{document}

\title{On photon angular momentum: transversality condition, Berry degree of freedom, and non-commutativity of photon position}

\author{Chun-Fang Li\footnote{Email address: cfli@shu.edu.cn}}

\affiliation{Department of Physics, Shanghai University, 99 Shangda Road, 200444
Shanghai, China}

\date{\today}

\begin{abstract}

Different from the usual conclusion that the separation of the photon angular momentum
into orbital and spin parts is physically meaningless, the orbital and spin angular momenta
are demonstrated in the first-quantization framework not to satisfy the standard commutation relation.
It is shown on the basis of the transversality condition that the spin is aligned with the propagation direction so that only the helicity can be the intrinsic degree of freedom. It is also shown on the same basis that only in the so-called intrinsic reference system does the helicity behave intrinsic.
The intrinsic reference system of the photon is determined by the ``action'' of a gauge potential, the Berry potential, on the helicity of the photon. The Berry potential is fixed by a vector-valued degree of freedom, called the Berry degree of freedom.
Because only the position of the photon in its intrinsic reference system is canonically conjugate to the momentum, the intrinsic reference system itself is endowed with quantum effects that depend on the Berry degree of freedom as well as the helicity.
This not only explains the so-called spin Hall effect of light but also helps to understand why the total angular momentum cannot be generally split into helicity-independent orbital and helicity-dependent spin parts.

\end{abstract}

\pacs{42.50.Tx, 03.65.Ca, 42.90.+m}
\maketitle


\section{Introduction}

The difficulty in understanding the physical properties of photon angular momentum is rooted in the constraint of transversality condition on the photon wavefunction. Usually it is argued \cite{Akhiezer, Bere, Cohen, Barnett-A, Barnett} on the basis of transversality condition that the separation of photon angular momentum into orbital and spin parts is physically meaningless. A representative argument \cite{Cohen} is as follows. The operator for the orbital angular momentum (OAM),
$\hat{\mathbf L}=\hat{\mathbf X} \times \hat{\mathbf P}$,
generates spatial rotations. But such operations cannot preserve the transversality of the photon wavefunction. The same is true of the spin. So the separation of the OAM from the spin has no physical meaning.
Nevertheless, by assuming that the OAM operator generates spatial rotations, it is meant \cite{EN, Enk, Bliokh} that the OAM operator satisfies the standard commutation relation,
\begin{equation}\label{FCR-L's}
[\hat{L}_i, \hat{L}_j] =i \hbar \epsilon_{ijk} \hat{L}_k,
\end{equation}
where $\epsilon_{ijk}$ is the Levi-Civit\'{a} pseudotensor.
This further requires \cite{Sakurai} that the position and momentum operators,
$\hat{\mathbf X}$ and $\hat{\mathbf P}$,
satisfy the following canonical commutation relations,
\begin{equation}\label{FCR}
[\hat{X}_i,\hat{X}_j] =0, \quad
[\hat{P}_i,\hat{P}_j] =0, \quad
[\hat{X}_i,\hat{P}_j] =i \hbar \delta_{ij}.
\end{equation}
This observation indicates that the constraint of transversality condition on the photon wavefunction does not necessarily rule out the separation of photon angular momentum into orbital and spin parts. Instead, it may imply that the OAM operator does not satisfy the standard commutation relation (\ref{FCR-L's}). In other words, it may imply that the position and momentum operators do not satisfy the canonical commutation relations (\ref{FCR}). The purpose of present paper is to show that this is indeed the case.
However, the commutation relations (\ref{FCR}) are so important that they were called by Dirac \cite{Dirac} the ``fundamental quantum conditions'' and were regarded by Sakurai \cite{Sakurai} as the ``cornerstone'' of quantum mechanics. Only quantities $\hat{\mathbf X}$ and $\hat{\mathbf P}$ that satisfy these commutation relations are said to be canonically conjugate to each other \cite{Cohen}.
If the photon position is not canonically conjugate to its momentum, what is the quantity that is canonically conjugate to the momentum? This is another important issue that we are concerned with.

The key point is that the position operator $\hat{\mathbf X}$ acting on the constrained wavefunction cannot be commutative,
$[\hat{X}_i,\hat{X}_j] \neq 0$,
though the momentum operator is,
$[\hat{P}_i,\hat{P}_j] =0$.
This is in agreement with the nonlocality of the photon in position space \cite{Jauch-P, Amrein, Pauli, Rosewarne-S}, which makes it impossible to consistently introduce a position operator with commuting components \cite{Pryce, Newton, Wightman, Jordan, Mourad, Pike, Hawton, Hawton-B} and to define a position-space wavefunction \cite{Akhiezer, Sipe, Bialynicki1996, Hawton99-1} in the usual sense \cite{Sakurai}.
Particularly, the transversality condition on the momentum-space ($\mathbf k$-space) wavefunction implies a gauge potential, the Berry potential \cite{Berry84}, which is fixed by a vector-valued degree of freedom, called the Berry degree of freedom.
It is nothing but the constant unit vector that was frequently introduced in the literature \cite{Stratton, Green, Pattanayak, Davis} and was extensively discussed in Refs. \cite{Li2008, Li09-2, Li09-1, Wang, Yang, Li-Y}.
What is intriguing is that for the helicity of the photon the Berry potential plays the role of some ``external'' field. Its ``action'' on the helicity of the photon determines the intrinsic reference system (IRS) of the photon. The IRS is therefore dependent on the Berry degree of freedom as well as the helicity.
Indeed, the Berry potential corresponds to a ``magnetic monopole'' \cite{Dirac31} of unit strength in $\mathbf k$-space \cite{Bialynicki1987, Fang}, with the Berry degree of freedom denoting the ``location'' of the singular line.
The significance to introduce the notion of IRS lies in the observation that the position of the photon in the IRS is canonically conjugate to the momentum. The canonical quantum numbers that follow thus describe only the properties of the photon relative to its IRS.
In addition, the transversality condition on the wavefunction implies that the spin of the photon lies entirely along the momentum direction \cite{Jauch, Lenstra, Mandel}. So the spin does not either satisfy the standard commutation relation,
\begin{equation}\label{CCR-S}
    [\hat{S}_i, \hat{S}_j] =i \hbar \epsilon_{ijk} \hat{S}_k,
\end{equation}
as is usually assumed \cite{EN, Enk, Bliokh}.
Especially, the magnitude of the spin, the helicity, behaves intrinsic only in the IRS.
As a result, the Berry degree of freedom, the unique ``external'' parameter to characterize the IRS, has observable physical effects that depend on the helicity.

These results are achieved as follows. The transversality condition is converted into a quasi unitary matrix that makes the vector wavefunction expressed in terms of a two-component wavefunction. Advantageous over the vector wavefunction, the two-component wavefunction is free of any constraints. But the transversality condition cannot solely determine the quasi unitary matrix. The constant unit vector that is introduced to completely determine the quasi unitary matrix turns out to be the Berry degree of freedom.
Unlike the vector wavefunction that is supposed to be defined over the laboratory reference system (LRS), the two-component wavefunction is shown to be defined over the IRS.
With the change of the Berry degree of freedom, one can relate a transformation on the two-component wavefunction without changing the vector wavefunction. That transformation is similar to the classical gauge transformation on the electromagnetic potentials in the sense that it does not change the electric and magnetic vectors of a radiation field.
However, with the change of the Berry degree of freedom, one can relate as well a transformation on the vector wavefunction without changing the two-component wavefunction. This transformation does not have classical counterpart, in that it means changing the electric and magnetic vectors. It is such a change of the Berry degree of freedom that is physically observable.
For clarity, the former transformation will be referred to as the Berry transformation of the first class and the latter one the Berry transformation of the second class.

It is clear that what we do in this paper is to uncover the physics that is hidden beyond the transversality condition from the quantum-mechanical point of view. Besides, we will discuss some related results that littered the literature, including
(a) the Pryce position operator \cite{Pryce}, which was also derived upon giving up the assumption of commuting components \cite{Mourad} and was used to discuss the properties of photon angular momentum \cite{Bliokh};
(b) the covariant derivative that was introduced by Bialynicki-Birula and Bialynicka-Birula \cite{Bialynicki1996, Bialynicki1975, Bialynicki1987, Bialynicki} in some two-component representation;
(c) the observation by Berry \cite{Berry98} that the spin does not ``give a complete description of the state of polarization'';
(d) and the complete set of eigenfunctions that van Enk and Nienhuis used in their second-quantization framework \cite{Enk}.
The contents are arranged as follows.

It is shown in Section \ref{TC-representation} that the 3-by-2 matrix introduced previously \cite{Li2008, Li09-2, Li09-1} in accordance with the transversality condition performs a quasi unitary transformation. It transforms the representation of vector wavefunctions into a representation of two-component wavefunctions. The constant unit vector that is introduced to determine the transformation matrix turns out to be the degree of freedom to specify the two-component representation.
The quasi unitary transformation is applied in Section \ref{spin} to explore the properties of photon spin. It is derived consistently for the first time in the first-quantization framework that the spin lies entirely along the momentum direction and therefore has commuting components, the same as van Enk and Nienhuis \cite{EN, Enk} found in a second-quantization framework.

The quasi unitary transformation is applied in Section \ref{intrinsic} to introduce the notion of IRS. It is shown that the position of the photon in the IRS is canonically conjugate to the momentum. The origin of the IRS, a constant of motion, conveys the ``action'' of the Berry potential on the helicity of the photon as if the Berry potential is an ``external'' field. The degree of freedom to specify the two-component representation is the Berry degree of freedom.
As a consequence, the position of the photon in the LRS is not commutative. It is also shown that the position in the LRS is invariant under the Berry transformation of the first class.
The IRS is demonstrated in Section \ref{observable} to have observable quantum effects.
First of all, it is shown that the two-component wavefunction is defined over the IRS so that any two-component representation is an intrinsic representation. This is to be contrasted with the vector representation, which is supposed to be a laboratory representation.
Secondly, it is found that any complete orthogonal set of eigenfunctions in the vector representation is always associated with one particular value of the Berry degree of freedom, demonstrating that the Berry degree of freedom is the unique ``external'' parameter to characterize the IRS.
So the Berry transformation of the second class does not have classical counterpart. It expresses some real physical process that substantially changes the IRS without affecting the position of the photon in the IRS.
For an eigen state of the helicity, it gives rise to a Berry phase.

The properties of photon OAM is explored in Section \ref{OAM}. The OAM of the photon about the origin of the LRS is split into the OAM of the photon concentrated at the origin of the IRS and the OAM of the photon about the origin of the IRS.
Because the origin of the IRS is dependent on the helicity, the total OAM cannot be independent of it \cite{Li09-1}. This explains why the total angular momentum of a light beam cannot be generally separated into helicity-independent OAM and helicity-dependent spin \cite{Barnett-A}.
On this basis, the commutation relation of the OAM that was obtained by Enk and Nienhuis \cite{EN, Enk} in a second-quantization framework is derived consistently for the first time in the first-quantization framework.
Section \ref{remarks} concludes the paper with remarks.

\section{From transversality condition to two-component representation}\label{TC-representation}

As is known \cite{Akhiezer, Cohen}, the $\mathbf k$-space wavefunction
$\mathbf{f} (\mathbf{k}, t)$
of the photon satisfies the Schr\"{o}dinger equation
\begin{equation}\label{Schrodinger-eq-L}
    i \frac{\partial \mathbf{f}}{\partial t}= \omega \mathbf{f}
\end{equation}
and is constrained by the transversality condition
\begin{equation}\label{TC}
    \mathbf{f}^\dag \mathbf{k}=0,
\end{equation}
where the angular frequency $\omega =ck$ plays the role of Hamiltonian, $k=|\mathbf{k}|$, and the superscript $\dag$ stands for the conjugate transpose.
Here vectors of three components, such as $\mathbf f$ and $\mathbf k$, are considered to be column matrices so that their scalar products are expressed as matrix multiplications.
Schr\"{o}dinger equation (\ref{Schrodinger-eq-L}) together with transversality condition (\ref{TC}) is strictly equivalent \cite{Akhiezer, Cohen} to the free-space Maxwell's equations.
This is because the electric and magnetic vectors of a radiation field that solve the free-space Maxwell's equations are uniquely determined by the vector wavefunction $\mathbf f$ via
\begin{subequations}\label{E-and-H}
\begin{align}
  \mathbf{E} (\mathbf{X},t) & =\frac{1}{(2 \pi)^{3/2}}
                               \int \sqrt{\frac{\hbar \omega}{2 \varepsilon_0}} \mathbf{f}
                                    \exp(i \mathbf{k} \cdot \mathbf{X}) d^3 k +c.c., \label{E-vec}\\
  \mathbf{H} (\mathbf{X},t) & =\frac{1}{(2 \pi)^{3/2}}
                               \int \sqrt{\frac{\hbar \omega}{2 \mu_0}} \mathbf{w} \times \mathbf{f}
                                    \exp(i \mathbf{k} \cdot \mathbf{X}) d^3 k +c.c.,
\end{align}
\end{subequations}
respectively, where $\mathbf{w}= \mathbf{k}/k$ is the unit wavevector.
In view of this, each vector wavefunction stands for a particular radiation field.
But the quantum mechanics \cite{Sakurai, Dirac} that we are familiar with assumes such a formalism in which the wavefunction is free of any constraints.
In order to explore what the transversality condition (\ref{TC}) implies in quantum mechanics, let us utilize it to transform the vector wavefunction $\mathbf f$ into a wavefunction that is not subject to any constraints and observe what follows in such a process.

\subsection{From transversality condition to quasi unitary transformation}

It is well known that the transversality condition (\ref{TC}) allows to expand the vector wavefunction $\mathbf f$ in terms of two orthogonal polarization vectors with respect to the wavevector.
Specifically, letting be $\mathbf u$ and $\mathbf v$ the two linear-polarization vectors that form with $\mathbf w$ a local Cartesian system \cite{Bialynicki1975, Mandel},
\begin{equation}\label{triad}
    \mathbf{u} \times \mathbf{v} =\mathbf{w}, \quad
    \mathbf{v} \times \mathbf{w} =\mathbf{u}, \quad
    \mathbf{w} \times \mathbf{u} =\mathbf{v},
\end{equation}
we can expand $\mathbf f$ as
$\mathbf{f} =f_1 \mathbf{u} +f_2 \mathbf{v}$.
Putting $\mathbf u$ and $\mathbf v$ together to construct a 3-by-2 matrix
$\varpi=(
          \begin{array}{cc}
            \mathbf{u} & \mathbf{v} \\
          \end{array}
        )
$,
we convert this equation into \cite{Li2008, Li09-2, Li09-1}
\begin{equation}\label{QUT-2}
    \mathbf{f} =\varpi \tilde{f},
\end{equation}
where
$
    \tilde{f} (\mathbf{k}, t)
         =\bigg(
            \begin{array}{c}
              f_1 \\
              f_2 \\
            \end{array}
          \bigg)
$
is a two-component quantity.
Of course, one can choose any two orthogonal polarization vectors to expand $\mathbf f$. Especially, one can choose the following two circular-polarization vectors,
\begin{equation*}
    \mathbf{c}_1 =\frac{1}{\sqrt 2} (\mathbf{u}+i \mathbf{v}), \quad
    \mathbf{c}_2 =\frac{1}{\sqrt 2} (\mathbf{u}-i \mathbf{v}).
\end{equation*}
In that case the resultant matrix
$\varpi^c=(
           \begin{array}{cc}
             \mathbf{c}_1 & \mathbf{c}_2 \\
           \end{array}
          )
$
is complex-valued. To keep consistent with previous expressions \cite{Li09-1}, we adopt here the real-valued matrix $\varpi$.

The matrix $\varpi$ in Eq. (\ref{QUT-2}) performs a quasi unitary transformation in the following sense.
On one hand, Eq. (\ref{QUT-2}) says that the matrix $\varpi$ acts on a two-component quantity $\tilde f$ and yields a vector wavefunction $\mathbf f$ satisfying the transversality condition (\ref{TC}). It is not difficult to show that
\begin{equation}\label{unitarity-2}
    \varpi^{\dag} \varpi =I_2,
\end{equation}
where $I_2$ is the 2-by-2 unit matrix.
On the other hand, multiplying both sides of Eq. (\ref{QUT-2}) by $\varpi^{\dag}$ from the left and considering Eq. (\ref{unitarity-2}), one gets
\begin{equation}\label{QUT-1}
    \tilde{f} =\varpi^{\dag} \mathbf{f}.
\end{equation}
It says that the matrix $\varpi^{\dag}$ acts on a vector wavefunction and yields a two-component quantity. A straightforward calculation gives
$\varpi \varpi^{\dag} =I_3-\mathbf{w} \mathbf{w}^{\dag}$,
where $I_3$ is the 3-by-3 unit matrix. But with the help of transversality condition (\ref{TC}), one has
\begin{equation*}
    \varpi \varpi^{\dag} \mathbf{f}=\mathbf{f}.
\end{equation*}
Keeping in mind that $\varpi^{\dag}$ and, therefore, $\varpi \varpi^{\dag}$ always act on the vector wavefunction $\mathbf f$ as is clearly shown in Eq. (\ref{QUT-1}), one may rewrite it simply as
\begin{equation}\label{unitarity-1}
    \varpi \varpi^{\dag} =I_3.
\end{equation}
Eqs. (\ref{unitarity-2}) and (\ref{unitarity-1}) express the quasi unitarity \cite{Golub} of the transformation matrix $\varpi$. $\varpi^{\dag}$ is the Moore-Penrose pseudo inverse of $\varpi$, and vice versa.

The quasi unitary transformation (\ref{QUT-2}) or (\ref{QUT-1}) establishes a one-to-one correspondence between the two-component quantity $\tilde f$ and the vector wavefunction $\mathbf f$. Furthermore, it makes the norm of $\tilde f$ the same as that of $\mathbf f$,
$\tilde{f}^{\dag} \tilde{f} =\mathbf{f}^{\dag} \mathbf{f}$. This means that the two-component quantity acts as a different kind of wavefunction and therefore constitutes a representation that is different from the vector representation.

\subsection{The degree of freedom to specify the two-component representation}\label{introducingI}

We have succeeded in converting the transversality condition (\ref{TC}) into a quasi unitary matrix $\varpi$ that is composed of the two orthogonal polarization vectors satisfying Eqs. (\ref{triad}). Its conjugate transpose transforms the vector representation into a two-component representation via Eq. (\ref{QUT-1}).
But the transversality condition (\ref{TC}) cannot solely determine $\varpi$; for Eqs. (\ref{triad}) cannot completely determine the polarization vectors up to a rotation about the wavevector \cite{Bialynicki1975, Mandel}. That is to say, the transversality condition cannot solely determine the two-component representation.
Let us point out \cite{Li2008, Li09-2, Li09-1} that one can introduce a constant unit vector, denoted here by $\mathbf I$, to completely determine the polarization vectors in the following way \cite{Green},
\begin{equation}\label{basis}
    \mathbf{u}_{\mathbf I} =\mathbf{v}_{\mathbf I} \times \frac{\mathbf k}{k},             \quad
    \mathbf{v}_{\mathbf I} =\frac{\mathbf{I} \times \mathbf{k}} {|\mathbf{I} \times \mathbf{k}|}.
\end{equation}
Indeed, the unit vectors $\mathbf{u}_{\mathbf I}$ and $\mathbf{v}_{\mathbf I}$ defined this way satisfy Eqs. (\ref{triad}).
The quasi unitary matrix composed of so defined polarization vectors is thus dependent on $\mathbf I$,
\begin{equation}\label{varpi}
\varpi_{\mathbf I}=(
                    \begin{array}{cc}
                      \mathbf{u}_{\mathbf I} & \mathbf{v}_{\mathbf I} \\
                    \end{array}
                   ).
\end{equation}
With $\varpi_\mathbf{I}$ we rewrite Eqs. (\ref{unitarity-2}) and (\ref{unitarity-1}) as
\begin{subequations}\label{unitarity}
\begin{align}
  \varpi^{\dag}_{\mathbf I} \varpi_{\mathbf I}        & = I_2, \label{unitarity-I2} \\
  \varpi_{\mathbf I}        \varpi^{\dag}_{\mathbf I} & = I_3, \label{unitarity-I1}
\end{align}
\end{subequations}
respectively.
The dependence of the matrix $\varpi_{\mathbf I}$ on $\mathbf I$ means that the two-component wavefunction for a given vector wavefunction is also dependent on $\mathbf I$. To reflect this, we rewrite Eqs. (\ref{QUT-2}) and (\ref{QUT-1}) explicitly as
\begin{subequations}\label{QUT-I}
\begin{align}
  \mathbf{f}            & =\varpi_{\mathbf I} \tilde{f}_{\mathbf I},  \label{f-bf}  \\
  \tilde{f}_{\mathbf I} & =\varpi^{\dag}_{\mathbf I} \mathbf{f},      \label{tildef-I}
\end{align}
\end{subequations}
respectively.
Now that the matrix (\ref{varpi}) is completely determined by the unit vector $\mathbf I$ via Eqs. (\ref{basis}), we can say that it together with Eqs. (\ref{basis}) is equivalent to the transversality condition (\ref{TC}) by virtue of transformation equation (\ref{f-bf}) or (\ref{tildef-I}).
Multiplying both sides of Eq. (\ref{Schrodinger-eq-L}) by $\varpi^{\dag}_{\mathbf I}$ from the left and making use of Eqs. (\ref{unitarity-I1}) and (\ref{tildef-I}), one arrives at the following Schr\"{o}dinger equation for the two-component wavefunction,
\begin{equation}\label{Sch-eq-tildef}
    i \frac{\partial \tilde{f}_{\mathbf I}}{\partial t}= \omega \tilde{f}_{\mathbf I},
\end{equation}
where the Hamiltonian is invariant under the transformation,
$\varpi^{\dag}_{\mathbf I} \omega \varpi_{\mathbf I}= \omega$.

What deserves noting is that there are no restrictions on the constant unit vector $\mathbf I$ in Eqs. (\ref{basis}). It can be represented by any point on the surface of unit sphere. That is to say, the unit vector to determine the transformation matrix $\varpi_{\mathbf I}$ is in reality the degree of freedom to specify the two-component representation.
Eq. (\ref{f-bf}) thus states that any given vector wavefunction can be expressed in terms of different two-component wavefunctions.
Let us see how this degree of freedom determines the two-component wavefunction for a given vector wavefunction $\mathbf f$.

Suppose that the unit vector $\mathbf I$ is changed into a different one, say $\mathbf{I}'$, so that the two-component wavefunction for the same vector wavefunction is given by
\begin{equation}\label{tildef'}
    \tilde{f}_{\mathbf{I}'}=\varpi^{\dag}_{\mathbf{I}'} \mathbf{f},
\end{equation}
where
$
\varpi_{\mathbf{I}'}=(\begin{array}{cc}
                        \mathbf{u}_{\mathbf{I}'} & \mathbf{v}_{\mathbf{I}'}
                      \end{array}
                     )
$
and
\begin{equation*}
    \mathbf{u}_{\mathbf{I}'} =\mathbf{v}_{\mathbf{I}'} \times \frac{\mathbf k}{k},
         \hspace{5pt}
    \mathbf{v}_{\mathbf{I}'} =\frac{\mathbf{I}' \times \mathbf{k}}
         {|\mathbf{I}' \times \mathbf{k}|}.
\end{equation*}
As remarked earlier, the polarization vectors
$\mathbf{u}_{\mathbf{I}'}$ and $\mathbf{v}_{\mathbf{I}'}$
that make up the new transformation matrix $\varpi_{\mathbf{I}'}$ are related to the old polarization vectors $\mathbf{u}_{\mathbf I}$ and $\mathbf{v}_{\mathbf I}$ by a rotation about $\mathbf k$. Letting be $\phi(\mathbf k)$ the $\mathbf k$-dependent rotation angle, such a rotation can be expressed as \cite{Tung}
\begin{equation}\label{rotation-pi}
    \varpi_{\mathbf{I}'}=\varpi_{\mathbf I} \exp \left(-i \hat{\sigma} \phi \right),
\end{equation}
where
\begin{equation}\label{sigma}
    \hat{\sigma}=\bigg(
                   \begin{array}{cc}
                     0 & -i \\
                     i &  0 \\
                   \end{array}
                 \bigg)
\end{equation}
is one of the Pauli matrices.
Substituting Eq. (\ref{rotation-pi}) into Eq. (\ref{tildef'}) and making use of Eq. (\ref{tildef-I}), one has
\begin{equation}\label{GT-tildef}
    \tilde{f}_{\mathbf{I}'}=\exp \left(i \hat{\sigma} \phi \right) \tilde{f}_{\mathbf I}.
\end{equation}

This is the transformation on the two-component wavefunction under the change of the degree of freedom $\mathbf I$, without changing the vector wavefunction. It is similar to the classical gauge transformation on electromagnetic potentials \cite{Jackson} in the sense that it does not change the electric and magnetic vectors of a radiation field as Eqs. (\ref{E-and-H}) indicate. We will see in Section \ref{intrinsic} that it is the Berry transformation of the first class.
But the degree of freedom $\mathbf I$ is not to be confused with the classical gauge degree of freedom.
The generator of transformation (\ref{GT-tildef}) is the Pauli matrix $\hat \sigma$, which will turn out to be the helicity operator in the two-component representation; whereas the classical gauge transformation on electromagnetic potentials does not have such a generator. After all, the two-component wavefunction is not equivalent to the electromagnetic potentials.
Rather, it is the vector wavefunction that is equivalent to the vector potential in the Coulomb gauge. As is known, in the Coulomb gauge the scalar potential vanishes and the vector potential $\mathbf A(\mathbf{X},t)$ is fixed by
$\nabla \cdot \mathbf{A}=0$.
The electric and magnetic vectors of a radiation field are expressed in the Coulomb gauge as
\begin{equation*}
    \mathbf{E}=-\frac{\partial \mathbf{A}}{\partial t}, \quad
    \mathbf{H}= \frac{1}{\mu_0} \nabla \times \mathbf{A},
\end{equation*}
respectively.
A comparison with Eqs. (\ref{E-and-H}) suggests that the vector potential has the following integral over plane waves,
\begin{equation*}
    \mathbf{A}=\frac{1}{(2 \pi)^{3/2}} \int \frac{\mathbf{a}(\mathbf{k},t)}{\sqrt 2}
                                            \exp(i\mathbf{k} \cdot \mathbf{X}) d^3 k +c.c.,
\end{equation*}
where the complex-valued function
$\mathbf{a}(\mathbf{k},t)$
in $\mathbf k$-space is one-to-one corresponding to the vector wavefunction by \cite{Cohen}
\begin{equation*}
    \mathbf{a}=-i \sqrt{\frac{\hbar}{\varepsilon_0 \omega}} \mathbf{f}.
\end{equation*}

In a word, we have converted the transversality condition (\ref{TC}) into the transformation equation (\ref{f-bf}) or (\ref{tildef-I}) that changes the Schr\"{o}dinger equation (\ref{Schrodinger-eq-L}) for the vector wavefunction into the Schr\"{o}dinger equation (\ref{Sch-eq-tildef}) for the two-component wavefunction.
So Eq. (\ref{Sch-eq-tildef}) together with the transformation equations (\ref{QUT-I}) is also equivalent to the free-space Maxwell's equations. One can describe a quantum state of photon either in the vector representation or in the two-component representation.
Advantageous over the vector wavefunction in Eq. (\ref{Schrodinger-eq-L}), the two-component wavefunction in Eq. (\ref{Sch-eq-tildef}) is no longer constrained by any conditions. It is thus instructive to analyze the properties of photon angular momentum in the two-component representation.

\subsection{Angular momentum in the two-component representation}

It is well known \cite{Darwin, Akhiezer, Cohen, Li09-1} that the angular momentum of a radiation field,
$\mathbf{J}=\varepsilon_0 \mu_0 \int \mathbf{X} \times (\mathbf{E} \times \mathbf{H}) d^3 X$,
about the origin of the LRS can be expressed in terms of the vector wavefunction as
\begin{equation}\label{TAM}
    \mathbf{J}=\int \mathbf{f}^{\dag} (\hat{\mathbf L} +\hat{\mathbf S}) \mathbf{f} d^3 k,
\end{equation}
where
\begin{subequations}\label{LandS}
\begin{align}
  \hat{\mathbf L} & = -\hat{\mathbf P} \times \hat{\mathbf X}, \label{L} \\
  \hat{\mathbf S} & = \hbar \hat{\mathbf \Sigma}, \label{S}
\end{align}
\end{subequations}
are the OAM and spin operators, respectively,
\begin{subequations}\label{XandP}
\begin{align}
  \hat{\mathbf P} &= \hbar \mathbf{k} \otimes I_3,    \label{P}\\
  \hat{\mathbf X} &= i \nabla_\mathbf{k} \otimes I_3, \label{X}
\end{align}
\end{subequations}
$\nabla_{\mathbf k}$ is the gradient operator with respect to $\mathbf k$, and
$(\hat{\Sigma}_k)_{ij} =-i \epsilon_{ijk}$.
The operators in Eqs. (\ref{XandP}) are indicated explicitly by $I_3$ to act on the vector wavefunction of three components.
Substituting Eq. (\ref{f-bf}) into Eq. (\ref{TAM}), one has
\begin{equation}\label{TAM-1}
    \mathbf{J}=\int \tilde{f}^{\dag}_\mathbf{I}
               (\hat{\mathbf l}_\mathbf{I} +\hat{\mathbf s}_\mathbf{I}) \tilde{f}_\mathbf{I} d^3 k,
\end{equation}
where
\begin{subequations}\label{lIandsI}
\begin{align}
  \hat{\mathbf l}_\mathbf{I} & =\varpi^{\dag}_\mathbf{I} \hat{\mathbf L} \varpi_\mathbf{I}
                               =-\hat{\mathbf p} \times \hat{\mathbf x}_\mathbf{I}
  \label{lI} \\
  \hat{\mathbf s}_\mathbf{I} & =\varpi^{\dag}_\mathbf{I} \hat{\mathbf S} \varpi_\mathbf{I}
  \label{sI}
\end{align}
\end{subequations}
are the OAM and spin operators, respectively, in the two-component representation,
\begin{subequations}\label{pandx}
\begin{align}
  \hat{\mathbf p}            & =\varpi^{\dag}_\mathbf{I} \hat{\mathbf P} \varpi_\mathbf{I}
                               =\hbar \mathbf{k} \otimes I_2                      \label{p} \\
  \hat{\mathbf x}_\mathbf{I} & =\varpi^{\dag}_\mathbf{I} \hat{\mathbf X} \varpi_\mathbf{I}
                               =\hat{\boldsymbol \xi} +\hat{\mathbf b}_\mathbf{I}
                                \label{xI}
\end{align}
\end{subequations}
are the momentum and position operators, respectively, and
\begin{subequations}\label{xiIandbI}
\begin{align}
  \hat{\boldsymbol \xi}        & =i \nabla_\mathbf{k} \otimes I_2,          \label{xi} \\
  \hat{\mathbf b}_\mathbf{I}   & =i \varpi^{\dag}_\mathbf{I} (\nabla_\mathbf{k} \varpi_\mathbf{I}). \label{bI}
\end{align}
\end{subequations}
The operator $\hat{\mathbf p}$ in Eq. (\ref{p}) and the operator $\hat{\boldsymbol \xi}$ in Eq. (\ref{xi}) are indicated explicitly by $I_2$ to act on the two-component wavefunction. They are independent of the degree of freedom $\mathbf I$.
The deep implication of their independence of $\mathbf I$ will be examined in Section \ref{observable}.

\section{The spin is not the generator of spatial rotations}\label{spin}

Let us first make use of the quasi unitary transformation (\ref{sI}) to explore the properties of the spin.
Substituting Eqs. (\ref{varpi}) and (\ref{S}) into Eq. (\ref{sI}) and decomposing the vector operator $\hat{\boldsymbol \Sigma}$ in the local Cartesian system $\mathbf{uvw}$ as
\begin{equation*}
    \hat{\boldsymbol \Sigma}
       =(\hat{\boldsymbol \Sigma} \cdot \mathbf{u}_{\mathbf I}) \mathbf{u}_{\mathbf I}
       +(\hat{\boldsymbol \Sigma} \cdot \mathbf{v}_{\mathbf I}) \mathbf{v}_{\mathbf I}
       +(\hat{\boldsymbol \Sigma} \cdot \mathbf{w}) \mathbf{w},
\end{equation*}
one gets
\begin{equation*}
    \hat{\mathbf s}_\mathbf{I}= \hbar \mathbf{w} \otimes \hat{\sigma},
\end{equation*}
where
\begin{equation*}
    \hat{\sigma}=\varpi^{\dag}_\mathbf{I} (\hat{\mathbf \Sigma} \cdot \mathbf{w}) \varpi_\mathbf{I}
                =\bigg(
                   \begin{array}{cc}
                     0 & -i \\
                     i &  0 \\
                   \end{array}
                 \bigg).
\end{equation*}
Since $\hat{\sigma}$ is exactly the constant Pauli matrix (\ref{sigma}) that is independent of $\mathbf I$, $\hat{\mathbf s}_\mathbf{I}$ is also independent of $\mathbf I$. Due to this, we omit its subscript and rewrite the above expression as
\begin{equation}\label{sI-sigma}
    \hat{\mathbf s}= \hbar \mathbf{w} \otimes \hat{\sigma}.
\end{equation}

Taking Eq. (\ref{sI-sigma}) into account, the inverse transformation of Eq. (\ref{sI}) gives for the spin operator in the vector representation,
\begin{equation}\label{S-VR}
    \hat{\mathbf S}=\hbar \mathbf{w} \otimes \hat{\Sigma}_{\mathbf w},
\end{equation}
where
\begin{equation}\label{Sigmaw}
    \hat{\Sigma}_{\mathbf w}
   \equiv \hat{\boldsymbol \Sigma} \cdot \mathbf{w}
   =      \varpi_\mathbf{I} \hat{\sigma} \varpi^{\dag}_\mathbf{I}.
\end{equation}
It follows from Eqs. (\ref{S}) and (\ref{S-VR}) that in the vector representation one actually has
\begin{equation}\label{vanishing-C}
    \mathbf{w} \times \hat{\boldsymbol \Sigma}=0.
\end{equation}
As a consequence, the Cartesian components of the spin commute,
\begin{equation}\label{CR-S's}
    [\hat{S}_i, \hat{S}_j]=0.
\end{equation}
This is what van Enk and Nienhuis \cite{EN, Enk} obtained in a second-quantization framework. It is different from the standard commutation relation (\ref{CCR-S}), meaning that the spin operator (\ref{S-VR}) is not the generator of spatial rotations of the vector wavefunction.

Eq. (\ref{sI-sigma}) or (\ref{S-VR}) expresses a well-known property \cite{Jauch, Lenstra, Mandel} that the photon spin lies entirely along the wavevector direction.
Here we arrive at it simply by making use of the representation transformation.
This shows that such a property is hidden beyond the transversality condition (\ref{TC}).
As a matter of fact, if the transversality condition were not taken into account, direct algebraic calculations with Eq. (\ref{S}) would give
$\hat{\mathbf S}^2= 2 \hbar^2$. Nevertheless, it is seen from Eq. (\ref{S-VR}) that
$\hat{\mathbf S}^2=\hbar^2 (1-\mathbf{w} \mathbf{w}^{\dag})$,
which is actually
$\hat{\mathbf S}^2=\hbar^2$
because $\hat{\mathbf S}^2$ acts only on the transverse vector wavefunction. This is the same as what can be obtained from Eq. (\ref{sI-sigma}):
$\hat{\mathbf s}^2=\hbar^2$.

Now that the spin is aligned with the wavevector direction, only its magnitude, \emph{the helicity, can be the intrinsic degree of freedom}.
But it should be pointed out that the helicity operator in the two-component representation is different from the helicity operator in the vector representation. The former is the constant Pauli matrix (\ref{sigma}); whereas the latter given by Eq. (\ref{Sigmaw}) is dependent on the wavevector.
This means that the helicity does not always behave intrinsic as we will see below.

\section{Identifying helicity-dependent IRS}\label{intrinsic}

Next, we turn our attention to the OAM.
As we have seen, the form of momentum operator remains ``unchanged'' under the transformation from the vector representation to the two-component representation. The momentum operator (\ref{p}) in the two-component representation is thus commutative,
\begin{equation}\label{CR-p}
    [\hat{p}_i, \hat{p}_j] =0.
\end{equation}
So in present and next sections we will only investigate the properties of the position by making use of the quasi unitary transformation (\ref{xI}). The discussions of the OAM will be left until Section \ref{OAM}.

\subsection{Position in the IRS is canonical}

It is seen from Eq. (\ref{xI}) that the operator in the two-component representation for the position in the LRS splits into two parts.
The first part $\hat{\boldsymbol \xi}$ takes the same gradient form as $\hat{\mathbf X}$ does, but acts on the two-component wavefunction.
Considering that no conditions such as Eq. (\ref{TC}) exist for the two-component wavefunction, its Cartesian components commute,
\begin{equation}\label{CR-xi}
    [\hat{\xi}_i, \hat{\xi}_j] =0.
\end{equation}
Besides, it satisfies the following commutation relation with the momentum operator,
\begin{equation}\label{CR-xip}
    [\hat{\xi}_i,\hat{p}_j] =i \hbar \delta_{ij}.
\end{equation}
Eqs. (\ref{CR-p})-(\ref{CR-xip}) are nothing but the canonical commutation relations between $\hat{\boldsymbol \xi}$ and $\hat{\mathbf p}$, meaning that this part represents such a position that is canonically conjugate to the momentum, called the canonical position.
It is the position of the photon in ``its own reference system''~\cite{Bertrand}.
It is worth emphasizing that the canonical commutation relations (\ref{CR-p})-(\ref{CR-xip}) are independent of the degree of freedom $\mathbf I$ because the canonical variables $\hat{\boldsymbol \xi}$ and $\hat{\mathbf p}$ are independent of it.

The operator $\hat{\boldsymbol \xi}$ for the canonical position is independent of the helicity as Eq. (\ref{xi}) indicates. Quantum-mechanically, it illustrates an important fact that \emph{only in ``its own reference system'' does the helicity of the photon behave intrinsic}.
We emphasize this because in the LRS the helicity does not behave intrinsic as will be clear shortly. For this reason, we call ``its own reference system'' the IRS. That is, the canonical position is the position of the photon in the IRS.
In view of this, the second part $\hat{\mathbf b}_{\mathbf I}$ of the position operator (\ref{xI}) represents the position of the origin of the IRS or, briefly, the position of the IRS in the LRS.
It is solely determined by $\varpi_{\mathbf I}$. Straightforward calculations give
\begin{equation}\label{bI-sigma}
    \hat{\mathbf b}_{\mathbf I}=\mathbf{A}_\mathbf{I} \otimes \hat{\sigma},
\end{equation}
where
\begin{equation}\label{AI}
    \mathbf{A}_\mathbf{I}=\frac{\mathbf{I} \cdot \mathbf{k}}{k |\mathbf{I} \times \mathbf{k}|}
                          \mathbf{v}_{\mathbf I},
\end{equation}
and $\hat \sigma$ is the helicity operator (\ref{sigma}). Obviously, its Cartesian components commute,
\begin{equation}\label{CR-b}
    \hat{\mathbf b}_{\mathbf I} \times \hat{\mathbf b}_{\mathbf I}=0.
\end{equation}
Moreover, being commutative with the Hamiltonian,
\begin{equation}\label{CRb-omega}
    [\hat{\mathbf b}_{\mathbf I}, \omega]=0,
\end{equation}
it is a constant of motion. The reason is that it is transverse in the sense that it is perpendicular to the wavevector.

We see from the quantum-mechanical point of view that the IRS denoted by its origin is dependent on the helicity though the position in the IRS is not. Hence, the position (\ref{xI}) in the LRS itself cannot be independent of the helicity.
As the name suggests, an intrinsic degree of freedom should be independent of the extrinsic degrees of freedom, such as the momentum as well as the position.
Taking this into consideration, one has to conclude that \emph{the helicity does not behave intrinsic in the LRS}.
With the help of Eqs. (\ref{CR-xi}) and (\ref{CR-b}), it is easy to find
\begin{equation}\label{CR-xI}
    \hat{\mathbf x}_\mathbf{I} \times \hat{\mathbf x}_\mathbf{I}
   =i \nabla_\mathbf{k} \times \hat{\mathbf b}_\mathbf{I}
   =i \mathbf{H}_\mathbf{I} \otimes \hat{\sigma},
\end{equation}
where
\begin{equation}\label{H}
    \mathbf{H}_\mathbf{I}=\nabla_\mathbf{k} \times \mathbf{A}_\mathbf{I}
                         =-\frac{\mathbf w}{k^2}, \quad
    \mathbf{w} \neq \pm \mathbf{I}.
\end{equation}
Different from the position in the IRS, the position in the LRS is not commutative. In other words, the Cartesian components of the position in the LRS are not compatible observables. This helps us to understand why there is no probability density for the photon position in the LRS \cite{Pauli}.

\subsection{Berry potential to determine the IRS}

Eq. (\ref{CR-xI}) demonstrates that the presence of the IRS leads to the non-commutativity of the position in the LRS. But how do we understand the physical meaning of the IRS? To find the answer, we substitute Eq. (\ref{bI-sigma}) into Eq. (\ref{xI}) to give
\begin{equation}\label{xI-xi}
    \hat{\mathbf x}_\mathbf{I} =\hat{\boldsymbol \xi} +\mathbf{A}_\mathbf{I} \otimes \hat{\sigma}.
\end{equation}
According to Barut and Bracken \cite{Barut}, if $\hat{\mathbf x}_\mathbf{I}$ is regarded as the analog of the kinematical momentum of a charged particle in an external magnetic field and the canonical position $\hat{\boldsymbol \xi}$ is regarded as the analog of the canonical momentum, then the helicity $\hat \sigma$ of the photon can be regarded as the analog of the electric charge of the particle and the vector quantity $\mathbf{A}_\mathbf{I}$ can be regarded as the analog of the vector potential of the magnetic field.
That is, $\mathbf{A}_\mathbf{I}$ serves as the gauge potential of some ``external'' field. In this sense, the IRS of the photon is the result of the ``action'' of such a gauge potential on the helicity of the photon.
Let us show that the unit vector $\mathbf I$ in $\mathbf{A}_\mathbf{I}$ is exactly the gauge degree of freedom to fix the gauge potential.

According to Eq. (\ref{xI-xi}), in a two-component representation that is specified by a different unit vector, say $\mathbf{I}'$, the operator for the position in the LRS is given by
\begin{equation}\label{xI'}
    \hat{\mathbf x}_{\mathbf{I}'} =\hat{\boldsymbol \xi} +\hat{\mathbf b}_{\mathbf{I}'},
\end{equation}
where
\begin{equation}\label{bI'}
    \hat{\mathbf b}_{\mathbf{I}'}=\mathbf{A}_{\mathbf{I}'} \otimes \hat{\sigma}
\end{equation}
and
\begin{equation}\label{AI'}
    \mathbf{A}_{\mathbf{I}'}=\frac{\mathbf{I}' \cdot \mathbf{k}}{k |\mathbf{I}' \times \mathbf{k}|}
                          \mathbf{v}_{\mathbf{I}'}.
\end{equation}
On the other hand, from Eq. (\ref{bI}) it follows that
$\hat{\mathbf b}_{\mathbf{I}'}=i \varpi^\dag_{\mathbf{I}'}(\nabla_\mathbf{k} \varpi_{\mathbf{I}'})$.
With the help of Eqs. (\ref{rotation-pi}) and (\ref{unitarity-I2}), one gets
\begin{equation}\label{GT-b}
    \hat{\mathbf b}_{\mathbf{I}'}
   =\hat{\mathbf b}_{\mathbf I}+\nabla_{\mathbf k} \phi \otimes \hat{\sigma}.
\end{equation}
An inspection of Eqs. (\ref{bI'}) and (\ref{bI-sigma}) gives
\begin{equation}\label{GT-A}
    \mathbf{A}_{\mathbf{I}'} =\mathbf{A}_\mathbf{I}+\nabla_{\mathbf k} \phi,
\end{equation}
showing that the potential does undergo a gauge transformation under the change of unit vector $\mathbf I$, with $\phi$ the corresponding gauge function.
In other words, the degree of freedom to specify the two-component representation is just the gauge degree of freedom to fix the gauge potential.
In fact, as can be seen from Eqs. (\ref{CR-xI}) and (\ref{H}), $\mathbf{A}_\mathbf{I}$ is nothing but the Berry potential \cite{Berry84} that corresponds to a ``magnetic monopole'' \cite{Dirac31} of unit strength in $\mathbf k$-space \cite{Bialynicki1987, Fang}.
The gauge degree of freedom $\mathbf I$, called the Berry degree of freedom, indicates the ``location'' of the monopole's singular line. It is thus concluded that the IRS of the photon is governed by the ``location'' of the monopole's singular line.

Now that $\mathbf I$ is the Berry degree of freedom, Eq. (\ref{GT-tildef}) expresses a class of gauge transformation in association with the Berry-potential transformation (\ref{GT-A}), referred to as the Berry transformation of the first class.
It is analogous to the gauge transformation on the wavefunction of a charged particle in an external magnetic field \cite{Sakurai, Barut}.
Specifically, the position in the LRS is invariant under such a gauge transformation. Indeed, taking Eq. (\ref{GT-b}) into account, it is easy to show that
\begin{equation*}
    \tilde{f}^\dag_{\mathbf{I}'} \hat{\mathbf x}_{\mathbf{I}'} \tilde{f}_{\mathbf{I}'}
   =\tilde{f}^\dag_\mathbf{I}    \hat{\mathbf x}_\mathbf{I}    \tilde{f}_\mathbf{I},
\end{equation*}
which is just
$\mathbf{f}^\dag \hat{\mathbf X} \mathbf{f}$
by virtue of Eqs. (\ref{tildef-I}) and (\ref{xI}).

\section{Quantum effects of the IRS}\label{observable}

It is clear that the Berry degree of freedom characterizes the IRS of the photon. Because only the position in the IRS is canonically conjugate to the momentum, the Berry degree of freedom is endowed with observable physical effects through the IRS. This is the difference of the Berry degree of freedom from the classical gauge degree of freedom. Let us show this below step by step.

\subsection{Intrinsic and laboratory representations}

\subsubsection{Distinguishing the intrinsic representation from the laboratory representation}

To this end, let us first discuss the physical meaning of the two-component wavefunction.
We have observed that the helicity operator (\ref{Sigmaw}) in the vector representation depends on the wavevector and that the helicity does not behave intrinsic in the LRS. This relation reflects that the wavefunction in the vector representation is defined over the LRS as is supposed.
But the wavefunction in the two-component representation is different. On one hand, the helicity operator in the two-component representation is the Pauli matrix (\ref{sigma}), which is obviously independent of the extrinsic degrees of freedom. On the other hand, as was just shown, the helicity behaves intrinsic in the IRS. Such a relation means that the wavefunction in the two-component representation is defined over the IRS. This is why the degree of freedom to specify the two-component representation is the degree of freedom to determine the IRS.
In particular, it follows \cite{Sakurai} from the canonical commutation relations (\ref{CR-p})-(\ref{CR-xip}) that the Fourier integral of the $\mathbf k$-space wavefunction $\tilde f$ in any two-component representation is the position-space wavefunction over the corresponding IRS,
\begin{equation*}
    \tilde{F} (\boldsymbol{\xi},t)
   =\frac{1}{(2\pi)^{3/2}}\int \tilde{f} (\mathbf{k},t)
    \exp(i \mathbf{k} \cdot \boldsymbol{\xi}) d^3 k.
\end{equation*}
By this it is meant that the momentum operator $\hat{\mathbf p}$ in the two-component representation is the generator of spatial translations in the IRS.
It is worth noting that the two-component representation specified by the Berry degree of freedom cannot be put in a relativistically covariant form. The reason is evident. As the canonical position, the position in the IRS is the generator of Galilean transformations \cite{Gottfried}, that is, the generator of momentum translations, rather than the generator of Lorentz transformations.
However, it is necessary to be aware that this does not mean the violation of relativistic invariance. The agreement of what we obtain here with Einstein's theory of relativity is ensured by the free-space Maxwell's equations that the electric and magnetic vectors (\ref{E-and-H}) satisfy.

Let us have a look at the differences between the two-component and vector representations.
The two-component representation is specified by the Berry degree of freedom that determines the helicity-dependent IRS via Eq. (\ref{bI-sigma}). Therefore, the wavefunction in a two-component representation is defined over the corresponding IRS. The helicity operator is given by Eq. (\ref{sigma}). The momentum operator is given by Eq. (\ref{p}). And the operator for the position in the LRS is given by Eq. (\ref{xI-xi}).
In contrast, the wavefunction in the vector representation is defined over the LRS. The helicity operator is given by Eq. (\ref{Sigmaw}). The momentum operator is given by Eq. (\ref{P}). And the operator for the position in the LRS is given by Eq. (\ref{X}).
To distinguish these two kinds of representations from each other, the two-component representation will be called the intrinsic representation and the vector representation will be called the laboratory representation. Accordingly, the wavefunction in the former will be called the intrinsic  wavefunction; and the wavefunction in the latter will be called the laboratory wavefunction.

\subsubsection{Position operator in the laboratory representation}

What is known in the literature is the laboratory representation. Nevertheless, the transversality condition on the laboratory wavefunction makes the issue of position operator rather delicate.

If the transversality condition (\ref{TC}) on the laboratory wavefunction were not taken into account, the operator (\ref{X}) for the position in the LRS would satisfy the canonical commutation relations (\ref{FCR}) with the momentum. This is, of course, not the case.
In fact, the inverse transformation of Eq. (\ref{xI}) gives for the operator for the position in the LRS,
\begin{equation}\label{X-Xi}
    \hat{\mathbf X}=\hat{\mathbf \Xi}_\mathbf{I}+\hat{\mathbf B}_{\mathbf I},
\end{equation}
where
\begin{equation}\label{XiI}
    \hat{\mathbf \Xi}_\mathbf{I}=\varpi_\mathbf{I} \hat{\boldsymbol \xi} \varpi^\dag_\mathbf{I}
                                =\hat{\mathbf X}-\hat{\mathbf B}_\mathbf{I}
\end{equation}
is the operator for the position in the IRS,
\begin{equation}\label{BI}
    \hat{\mathbf B}_\mathbf{I}=\varpi_\mathbf{I} \hat{\mathbf b}_\mathbf{I} \varpi^{\dag}_\mathbf{I}
                              =-i \varpi_\mathbf{I} (\nabla_\mathbf{k} \varpi^{\dag}_\mathbf{I})
\end{equation}
represents the origin of the IRS and is given by
\begin{equation}\label{BI-Sigma}
    \hat{\mathbf B}_\mathbf{I}=\mathbf{A}_\mathbf{I} \otimes \hat{\Sigma}_{\mathbf w}
\end{equation}
by virtue of Eqs. (\ref{bI-sigma}) and (\ref{Sigmaw}).
Here we arrive again at the conclusion in the laboratory representation that the IRS results from the action of the Berry potential on the helicity.

The counterpart of Eq. (\ref{CR-xi}) in the laboratory representation assumes
$\hat{\mathbf \Xi}_\mathbf{I} \times \hat{\mathbf \Xi}_\mathbf{I}=0$;
and the counterpart of Eq. (\ref{CR-b}) reads
$\hat{\mathbf B}_{\mathbf I} \times \hat{\mathbf B}_{\mathbf I}=0$.
Since
\begin{equation}\label{GL}
    \nabla_\mathbf{k} \hat{\Sigma}_\mathbf{w}
   =\frac{(\mathbf{w} \times \hat{\mathbf \Sigma}) \times \mathbf{w}}{k}=0
\end{equation}
by virtue of Eq. (\ref{vanishing-C}), it follows from Eq. (\ref{X-Xi}) that
\begin{equation}\label{CR-X}
    \hat{\mathbf X} \times \hat{\mathbf X}
   =i \nabla_\mathbf{k} \times \hat{\mathbf B}_\mathbf{I}
   =i \mathbf{H}_\mathbf{I} \otimes \hat{\Sigma}_{\mathbf w},
\end{equation}
which is parallel to Eq. (\ref{CR-xI}) in the intrinsic representation.
In a word, although it takes the simple gradient form with respect to the wavevector in the laboratory representation, the operator $\hat{\mathbf X}$ for the position in the LRS is actually noncommutative. This can be explained as follows.
Usually, the transversality condition (\ref{TC}) is interpreted to mean that the three Cartesian components of the laboratory wavefunction are not independent of one another. This is because the three Cartesian components of the wavevector are supposed to be independent of one another.
But mathematically the transversality condition (\ref{TC}) is about the directional relation between two vectors. We can interpret it as well the other way around.
That is, if the Cartesian components of the laboratory wavefunction are regarded as independent, then the Cartesian components of the wavevector will not be independent.
Considering that the wavevector is the argument of the laboratory wavefunction, the partial derivatives of the laboratory wavefunction with respect to different Cartesian components of the wavevector cannot be independent. This is just what Eq. (\ref{CR-X}) shows.

Eq. (\ref{CR-X}) shows that the position in the LRS is not canonically conjugate to the momentum. As a result, the position operator $\hat{\mathbf X}$ is no longer the generator of momentum translations.
This is not difficult to understand. After all, photons are extremely relativistic particles, whereas the momentum translations are the Galilean transformations, which are inconsistent with Einstein's theory of relativity.
Due to the same reason, the momentum operator is not the generator of spatial translations in the LRS.
It is no wonder why the Fourier integral of the $\mathbf k$-space laboratory wavefunction $\mathbf f$ cannot be interpreted in the usual sense as the position-space wavefunction in the LRS \cite{Akhiezer, Bere}.

In addition, substituting Eqs. (\ref{X}) and (\ref{BI-Sigma}) into Eq. (\ref{XiI}), one gets
\begin{equation}\label{Xi-Sigma}
    \hat{\mathbf \Xi}_\mathbf{I}=i \nabla_\mathbf{k} \otimes I_3
                                -\mathbf{A}_\mathbf{I} \otimes \hat{\Sigma}_\mathbf{w}.
\end{equation}
On the other hand, substituting Eq. (\ref{xi}) into Eq. (\ref{xI-xi}), one has
\begin{equation}\label{xI-sigma}
    \hat{\mathbf x}_\mathbf{I} =i \nabla_\mathbf{k} \otimes I_2
                               +\mathbf{A}_\mathbf{I} \otimes \hat{\sigma}.
\end{equation}
These two expressions look rather similar. They depend on the Berry potential in almost the same way. But, as shown before, they have quite different physical meanings. The former is the operator for the position in the IRS in the laboratory representation, whereas the latter is the operator for the position in the LRS in the intrinsic representation.

Furthermore, it is pointed out that what is commonly called \cite{Mourad, Bliokh} the Pryce position operator \cite{Pryce} in the laboratory representation,
\begin{equation*}
    \hat{\mathbf X}_P= \hat{\mathbf X} +\frac{\mathbf{k} \times \hat{\mathbf S}}{k^2},
\end{equation*}
is actually the position operator $\hat{\mathbf X}$. This is because the spin $\hat{\mathbf S}$ is aligned with the wavevector direction so that
$\mathbf{k} \times \hat{\mathbf S}$ vanishes. Taking this into account, the so-called commutative position operator introduced by Hawton \cite{Hawton} for $\alpha=0$ is in reality the operator (\ref{Xi-Sigma}) for the position in the IRS in the case of  $\mathbf{I}=\mathbf{e}_z$.
Frankly speaking, Bialynicki-Birula and Bialynicka-Birula \cite{Bialynicki1996, Bialynicki1975, Bialynicki1987, Bialynicki} once introduced, in a representation of two-component wavefunctions, an expression that is similar to Eq. (\ref{xI-sigma}). They denoted it by $i \mathbf{D}_\mathbf{k}$ and called $\mathbf{D}_\mathbf{k}$ the covariant derivative \cite{Bialynicki1996}. But they failed to find the simple dependence of the Berry potential on the Berry degree of freedom $\mathbf I$.

\subsection{Complete sets of eigenfunctions in the laboratory representation}\label{eigenfunctions}

Now it is clear that the reason for us not to be able to easily find a complete set of eigenfunctions in the laboratory representation \cite{Enk} is that the position in the LRS is not canonically conjugate to the momentum.
Thanks to the constant helicity operator $\hat \sigma$ and the canonical commutation relations (\ref{CR-p})-(\ref{CR-xip}), it is straightforward to write out complete sets of eigenfunctions in the intrinsic representation.
Furthermore, with the help of quasi unitary transformation (\ref{f-bf}), we can conveniently convert such a complete set into a complete set of eigenfunctions in the laboratory representation. Let us show, by discussing complete sets of eigenfunctions in the laboratory representation, why the Berry degree of freedom has observable physical effects.

As is known, the canonical commutation relations (\ref{CR-p})-(\ref{CR-xip}) determine a maximal set of three compatible dynamical variables. Their eigenvalues give a complete set of canonical quantum numbers, denoted collectively by $q$.
Furthermore, from the constant helicity operator $\hat \sigma$ it follows that these quantum numbers together with the helicity quantum number $\sigma$ constitute a complete set of four quantum numbers.
Denoted by $\{ \tilde{f}_{\sigma q} \}$, a complete orthonormal set of eigenfunctions in the intrinsic representation takes the form,
\begin{equation}\label{EF}
    \tilde{f}_{\sigma q}=\tilde{\alpha}_\sigma f_q,
\end{equation}
where
\begin{equation*}
    \tilde{\alpha}_{\pm 1}=\frac{1}{\sqrt{2}} \bigg(
                                                \begin{array}{c}
                                                      1 \\
                                                  \pm i \\
                                                \end{array}
                                              \bigg)
\end{equation*}
are the eigenvectors of helicity operator $\hat \sigma$ with eigenvalues $\sigma= \pm 1$, satisfying
\begin{equation}\label{EV-eq}
    \hat{\sigma} \tilde{\alpha}_{\sigma}= \sigma \tilde{\alpha}_{\sigma},
\end{equation}
and $f_q$ denotes the simultaneous eigenfunction of a set of compatible dynamical variables.
The orthonormality relation for $\{ \tilde{f}_{\sigma q} \}$ assumes
\begin{equation}\label{orthonormality1}
    \int \tilde{f}^\dag_{\sigma' q'} \tilde{f}_{\sigma q} d^3 k
   =\delta_{\sigma' \sigma} \delta_{q' q},
\end{equation}
where the Kronecker $\delta_{q' q}$ should be replaced with the Dirac $\delta$-function for continuous canonical quantum numbers.
It is emphasized that the complete orthonormal set of eigenfunctions in the intrinsic representation does not depend on the Berry degree of freedom, though the intrinsic representation has to be specified by this degree of freedom. This is because the canonical commutation relations (\ref{CR-p})-(\ref{CR-xip}) as well as the helicity operator $\hat \sigma$ have nothing to do with it. In other words, both the canonical quantum numbers and the helicity quantum number describe only the quantum properties of the photon relative to its IRS.
The term ``IRS'' is indeed deemed appropriate.

Nevertheless, because the position in the LRS is not canonically conjugate to the momentum, we cannot find a complete orthonormal set of eigenfunctions for the laboratory representation in the way in which we do for the intrinsic representation.
But the quasi unitary transformation (\ref{f-bf}) allows us to readily obtain from
$\{ \tilde{f}_{\sigma q} \}$
a complete orthonormal set of eigenfunctions in the laboratory representation, denoted by
$\{ \mathbf{f}_{\mathbf{I}, \sigma q} \}$,
as follows,
\begin{equation}\label{f-EF}
    \mathbf{f}_{\mathbf{I}, \sigma q} =\varpi_\mathbf{I} \tilde{f}_{\sigma q}.
\end{equation}
The orthonormality relation for $\{ \mathbf{f}_{\mathbf{I}, \sigma, q} \}$ reads
\begin{equation}\label{orthonormality2}
    \int \mathbf{f}^\dag_{\mathbf{I}, \sigma' q'} \mathbf{f}_{\mathbf{I}, \sigma q} d^3 k
   =\delta_{\sigma' \sigma} \delta_{q' q},
\end{equation}
by virtue of Eqs. (\ref{unitarity-I2}) and (\ref{orthonormality1}).
This shows that to each complete orthonormal set of eigenfunctions in an intrinsic representation that is specified by a particular value of the Berry degree of freedom, there corresponds in the laboratory representation a complete orthonormal set of eigenfunctions that is dependent on the same value.

This is unexpected. Given the helicity and canonical quantum numbers, one still needs the Berry degree of freedom to determine an eigen state of the photon!
But this is understandable.
As we just mentioned, the helicity and canonical quantum numbers describe only the quantum properties of the photon relative to its IRS. In order to completely determine an eigen state of the photon, it is essential to know how to determine its IRS. This is done by the action of the Berry potential on the helicity of the photon.
We have seen that the Berry potential is an ``external'' field.
The complete orthonormal set of eigenfunctions that we just gave in Eq. (\ref{f-EF}) demonstrates further that different values of the Berry degree of freedom signify different ``external'' fields.
This shows that the Berry degree of freedom serves as the unique ``external'' parameter to characterize the ``external'' field.
That is to say, the Berry degree of freedom is the unique ``external'' parameter to characterize the IRS (\ref{BI-Sigma}).

Let us consider an important case in which the intrinsic wavefunction, $\tilde f$, of a photon state satisfies
\begin{equation}\label{zeromean}
    \int \tilde{f}^\dag \hat{\boldsymbol \xi} \tilde{f} d^3 k=0,
\end{equation}
so that the origin of the IRS reduces to the barycenter of the photon in the sense that
\begin{equation*}
    \int \tilde{f}^\dag \hat{\mathbf b}_\mathbf{I} \tilde{f} d^3 k
   =\int \tilde{f}^\dag \hat{\mathbf x}_\mathbf{I} \tilde{f} d^3 k.
\end{equation*}
In this case, the Berry degree of freedom is the ``external'' parameter to characterize the barycenter. One of the examples of eigen states having property (\ref{zeromean}) is the plane-wave state in position space. Its intrinsic wavefunction is given by
\begin{equation*}
    \tilde{f}_{\sigma \mathbf{k}_0}=\tilde{\alpha}_\sigma \delta^3 (\mathbf{k}-\mathbf{k}_0)
                                    \exp(-i \omega t),
\end{equation*}
where $\mathbf{k}_0$ denotes the eigen momentum. It is in fact the eigen state of barycenter operator $\hat{\mathbf b}_\mathbf{I}$ with eigenvalue:
\begin{equation*}
    \mathbf{b}_{\mathbf{I}, \sigma \mathbf{k}_0}
   =\sigma \frac{\mathbf{I} \cdot \mathbf{k}_0}{k_0 |\mathbf{I} \times \mathbf{k}_0|^2}
    (\mathbf{I} \times \mathbf{k}_0),
\end{equation*}
where $k_0= |\mathbf{k}_0|$.
Other examples include the diffraction-free beams in position space that were discussed in Ref. \cite{Yang} and the spherical waves that will be discussed in Section \ref{OAM}.

In a word, the eigen state of the photon in free space is dependent on the Berry degree of freedom, which appears as the ``external'' parameter to characterize the photon barycenter.
It is pointed out that the Berry degree of freedom in the eigenfunctions that were used in the second-quantization framework by van Enk and Nienhuis \cite{Enk} is along the $z$ axis \cite{Li-Y}. As a matter of fact, the equations (41) in Ref. \cite{Enk} are the non-normalized form of our equations (\ref{basis}) in the case of $\mathbf{I}=\mathbf{e}_z$.
Let us see further how the Berry degree of freedom affects the quantum state of the photon.

\subsection{Berry transformation of the second class}

Once the values of quantum numbers $\sigma$ and $q$ are given, the eigenfunction (\ref{EF}) is determined regardless of the concrete intrinsic representation. So we consider one given intrinsic wavefunction $\tilde f$.
To simplify our discussions, we will assume in the remainder of this section that the intrinsic wavefunction satisfies Eq. (\ref{zeromean}) so that the origin of the IRS reduces to the photon barycenter.
If the intrinsic wavefunction is defined over an IRS that is characterized by unit vector $\mathbf I$, the corresponding laboratory wavefunction is given by
\begin{equation}\label{f-I}
    \mathbf{f}_\mathbf{I}=\varpi_\mathbf{I} \tilde{f},
\end{equation}
in accordance with Eq. (\ref{f-bf}); and the corresponding barycenter is represented by operator (\ref{BI-Sigma}) in the laboratory representation.
If the intrinsic wavefunction is defined over a different IRS that is characterized by unit vector, say $\mathbf{I}'$, the corresponding laboratory wavefunction is given by
\begin{equation}\label{f-I'}
    \mathbf{f}_{\mathbf{I}'}=\varpi_{\mathbf{I}'} \tilde{f};
\end{equation}
and the corresponding barycenter is represented by
\begin{equation}\label{BI'}
    \hat{\mathbf B}_{\mathbf{I}'}=\mathbf{A}_{\mathbf{I}'} \otimes \hat{\Sigma}_{\mathbf w},
\end{equation}
where $\mathbf{A}_{\mathbf{I}'}$ is given by Eq. (\ref{AI'}).
But according to Eq. (\ref{BI}), one has
$\hat{\mathbf B}_{\mathbf{I}'}=-i \varpi_{\mathbf{I}'}
                                (\nabla_\mathbf{k} \varpi^{\dag}_{\mathbf{I}'})$.
Upon making use of Eqs. (\ref{rotation-pi}) and (\ref{Sigmaw}), one gets
\begin{equation}\label{GT-B}
    \hat{\mathbf B}_{\mathbf{I}'}=\hat{\mathbf B}_\mathbf{I}
                                 +(\nabla_\mathbf{k} \phi) \otimes \hat{\Sigma}_\mathbf{w}.
\end{equation}
Obviously, it also reflects the Berry-potential transformation (\ref{GT-A}).

From Eqs. (\ref{f-I}) and (\ref{unitarity-I2}) it follows that
$\tilde{f}=\varpi^\dag_\mathbf{I} \mathbf{f}$. Substituting it into Eq. (\ref{f-I'}) and taking Eqs. (\ref{rotation-pi}) and (\ref{Sigmaw}) into account, one has
\begin{equation}\label{GT-f}
    \mathbf{f}_{\mathbf{I}'}=\exp(-i \hat{\Sigma}_\mathbf{w} \phi) \mathbf{f}_\mathbf{I}.
\end{equation}
This is the transformation on the laboratory wavefunction under the change of the Berry degree of freedom, without changing the intrinsic wavefunction, referred to as the Berry transformation of the second class.
In much the same way as the Berry transformation of the first class is generated by the helicity operator $\hat \sigma$ in the intrinsic representation, it is generated by the helicity operator $\hat{\Sigma}_\mathbf{w}$ in the laboratory representation.
In addition, to the Berry transformation (\ref{GT-f}) there also corresponds an ``invariance''. It is the invariance of the position in the IRS.
Indeed, letting
$\hat{\mathbf \Xi}_{\mathbf{I}'}=\hat{\mathbf X}-\hat{\mathbf B}_{\mathbf{I}'}$
in accordance with Eq. (\ref{XiI}) and noticing Eq. (\ref{GL}), it is a straightforward calculation to show
$
    \mathbf{f}^\dag_{\mathbf{I}'} (\hat{\mathbf \Xi}_{\mathbf{I}'}) \mathbf{f}_{\mathbf{I}'}
   =\mathbf{f}^\dag_\mathbf{I}    (\hat{\mathbf \Xi}_\mathbf{I})    \mathbf{f}_\mathbf{I},
$
which is just
$\tilde{f}^\dag \hat{\boldsymbol \xi} \tilde{f}$.
Nevertheless, the Berry transformation (\ref{GT-f}) corresponds to a substantial change in the photon barycenter as is explicitly shown by Eq. (\ref{GT-B}).
As a consequence, it changes the position of the photon in the LRS.
It deserves emphasizing that the barycenter of a free photon is a constant of motion. By this it is meant that a photon cannot change its barycenter unless undergoing real physical actions other than the action of the ``magnetic monopole'' in $\mathbf k$-space. In view of this, the Berry transformation (\ref{GT-f}) expresses some real physical action.

Although it changes the electric vector as Eq. (\ref{E-vec}) indicates, the Berry transformation (\ref{GT-f}) does not change the helicity,
$
    \mathbf{f}^\dag_{\mathbf{I}'} (\hat{\Sigma}_{\mathbf w}) \mathbf{f}_{\mathbf{I}'}
   =\mathbf{f}^\dag_\mathbf{I}    (\hat{\Sigma}_{\mathbf w}) \mathbf{f}_\mathbf{I},
$
and hence does not change the spin by virtue of Eq. (\ref{S-VR}). This explains why the spin is not able to completely describe the vector property (or ``the state of polarization'') of a radiation field \cite{Berry98}.

\subsection{Eigen states of helicity and Berry phase}

At the end of this section let us point out that the transformation factor
$\exp(-i \hat{\Sigma}_\mathbf{w} \phi)$
in Eq. (\ref{GT-f}) becomes a Berry phase factor for the eigen state of helicity.
To this end, it is only necessary to show that if an intrinsic wavefunction is the eigenfunction of helicity operator $\hat \sigma$, the corresponding laboratory wavefunction must be the eigenfunction of helicity operator $\hat{\Sigma}_\mathbf{w}$.
Let the intrinsic wavefunction $\tilde f$ be the eigenfunction of helicity operator $\hat \sigma$, having the form
$\tilde{f}_\sigma =\tilde{\alpha}_\sigma f$,
where $f$ is any physically allowed function. In this case, the laboratory wavefunction (\ref{f-I}) becomes
\begin{equation*}
    \mathbf{f}_{\mathbf{I}, \sigma} = \varpi_{\mathbf I}   \tilde{f}_{\sigma},
\end{equation*}
and Eq. (\ref{GT-f}) becomes
\begin{equation}\label{GT-f-Sigma}
    \mathbf{f}_{\mathbf{I}', \sigma}=\exp(-i \hat{\Sigma}_\mathbf{w} \phi)
                                     \mathbf{f}_{\mathbf{I}, \sigma}.
\end{equation}
It is straightforward to show by making use of Eqs. (\ref{Sigmaw}), (\ref{unitarity-I2}), and (\ref{EV-eq}) that
\begin{equation*}
    \hat{\Sigma}_\mathbf{w} \mathbf{f}_{\mathbf{I}, \sigma}
       =\sigma \mathbf{f}_{\mathbf{I}, \sigma},
\end{equation*}
meaning that $\mathbf{f}_{\mathbf{I}, \sigma}$ is indeed the eigenfunction of helicity operator
$\hat{\Sigma}_\mathbf{w}$.
With its help, Eq. (\ref{GT-f-Sigma}) further becomes
\begin{equation}\label{GT-f-sigma}
    \mathbf{f}_{\mathbf{I}', \sigma}=\exp(-i \sigma \phi) \mathbf{f}_{\mathbf{I}, \sigma}.
\end{equation}
Clearly, the change of $\mathbf I$ in this case makes the laboratory wavefunction acquire a phase. Corresponding to the Berry-potential transformation (\ref{GT-A}), this is a Berry phase.
It is dependent not only on the helicity quantum number but also on the wavevector. When $\mathbf{f}_{\mathbf{I}', \sigma}$ in Eq. (\ref{GT-f-sigma}) is substituted into Eqs. (\ref{E-and-H}), the phase will inevitably have its impact on the electric and magnetic vectors.
It is such an impact that changes the photon barycenter in accordance with Eq. (\ref{GT-B}).
A previous analysis \cite{Li09-2} showed that the refraction process at the interface between two dielectric media is a real physical action that changes the direction of $\mathbf I$ with respect to the propagation direction.
This is why the so-called spin Hall effect of light in such a process \cite{Hosten} can be explained in terms of the Berry phase as well \cite{Onoda}, even though the process is not adiabatic.

To conclude this section, we summarize in brief that the quantum effects of the Berry degree of freedom lies in the observation that the fundamental quantum conditions (\ref{CR-p})-(\ref{CR-xip}) hold only in the IRS. This is the primary result of present paper. Such effects are hidden beyond the transversality condition (\ref{TC}) on the laboratory wavefunction.

\section{Effects of the IRS on the OAM}\label{OAM}

We have seen in Section \ref{spin} how the transversality condition (\ref{TC}) on the laboratory wavefunction affects the properties of the spin. Now we are in a position to see how it affects the properties of the OAM.

\subsection{The OAM is dependent on the helicity}

The same as the position operator (\ref{xI}) in the intrinsic representation, the operator in the intrinsic representation for the OAM about the origin of the LRS also splits into two parts,
\begin{equation}\label{l}
    \hat{\mathbf l}_\mathbf{I}=\varpi^\dag_\mathbf{I} \hat{\mathbf L} \varpi_\mathbf{I}
                              =\hat{\boldsymbol \lambda} +\hat{\mathbf m}_\mathbf{I}.
\end{equation}
The first part
$\hat{\boldsymbol \lambda}=-\hat{\mathbf p} \times \hat{\boldsymbol \xi}$
is the OAM of the photon about the origin of the IRS. It is a constant of motion,
\begin{equation}\label{CR-lambda and omega}
    [\hat{\boldsymbol \lambda}, \omega] =0.
\end{equation}
Thanks to the fundamental quantum conditions (\ref{CR-p})-(\ref{CR-xip}), it satisfies the canonical commutation relation of the angular momentum \cite{Sakurai},
\begin{equation}\label{CR-lambda}
    [\hat{\lambda}_i, \hat{\lambda}_j] =i \hbar \epsilon_{ijk} \hat{\lambda}_k.
\end{equation}
The same as the position in the IRS, it is also independent of the Berry degree of freedom.
The second part
\begin{equation}\label{mI}
    \hat{\mathbf m}_{\mathbf I}
   \equiv \hat{\mathbf b}_{\mathbf I} \times \hat{\mathbf p}
   =\hbar \frac{\mathbf{I} \cdot \mathbf{k}}{|\mathbf{I} \times \mathbf{k}|}
    \mathbf{u}_{\mathbf I} \otimes \hat{\sigma}
\end{equation}
is the OAM of the photon concentrated at the origin of the IRS. Like $\hat{\boldsymbol \lambda}$, it is also a constant of motion,
\begin{equation}\label{CR-Lambda and omega}
    [\hat{\mathbf m}_{\mathbf I}, \omega] =0.
\end{equation}
But different from $\hat{\boldsymbol \lambda}$, its Cartesian components commute,
\begin{equation}\label{CR-m}
    \hat{\mathbf m}_{\mathbf I} \times \hat{\mathbf m}_{\mathbf I}=0.
\end{equation}
Besides, it is dependent on the Berry degree of freedom.

From Eqs. (\ref{CR-lambda and omega}) and (\ref{CR-Lambda and omega}) it follows that the total OAM is a constant of motion, too.
But it is important to note that the helicity dependence of the origin of the IRS makes the second part of the OAM depend also on the helicity. Because the first part does not depend on the helicity, the total OAM cannot be independent of the helicity \cite{Li09-1}. This extraordinary result explains why the total angular momentum of a non-paraxial beam cannot be separated into helicity-independent OAM and helicity-dependent spin \cite{Barnett-A, Li09-1}.

\subsection{The OAM is not the generator of spatial rotations}

It is a straightforward calculation to show that the first part of the OAM does not commute with the spin,
\begin{equation*}
    [\hat{\lambda}_i, \hat{s}_j]= i \hbar \epsilon_{ijk} \hat{s}_k,
\end{equation*}
though the second part does. As a result, the total OAM does not commute with the spin either,
\begin{equation}\label{CR-lands}
    [(\hat{l}_i)_\mathbf{I}, \hat{s}_j]= i \hbar \epsilon_{ijk} \hat{s}_k,
\end{equation}
in contrary to the usual assumption \cite{Akhiezer, Enk}.
With the help of Eqs. (\ref{CR-lambda}) and (\ref{CR-m}), it is not difficult to find
\begin{equation}\label{CR-OAM}
    [(\hat{l}_i)_\mathbf{I}, (\hat{l}_j)_\mathbf{I}]
   =i \hbar \epsilon_{ijk} \{ (\hat{l}_k)_\mathbf{I}-\hat{s}_k \}.
\end{equation}
The inverse of transformations (\ref{sI}) and (\ref{l}) leads to the counterpart in the laboratory representation
\begin{equation}\label{CR-L's}
    [\hat{L}_i, \hat{L}_j] =i \hbar \epsilon_{ijk} (\hat{L}_k -\hat{S}_k).
\end{equation}
This is the commutation relation of the OAM that was found by van Enk and Nienhuis \cite{EN, Enk}. Here we arrive at it without resorting to the second quantization. Clearly, Eq. (\ref{CR-L's}) is different from the standard commutation relation (\ref{FCR-L's}). The OAM operator $\hat{\mathbf L}$ is thus not the generator of spatial rotations as is usually assumed \cite{Cohen, Bliokh}.

From Eqs. (\ref{sI-sigma}), (\ref{l}), and (\ref{mI}) it follows that the operator for the total angular momentum in the intrinsic representation reads
\begin{equation}\label{TAM-BR}
    \hat{\mathbf j}_\mathbf{I}
   =\hat{\mathbf s} +\hat{\mathbf l}_\mathbf{I}
   =\hat{\boldsymbol \lambda} +\hbar \frac{\mathbf{I} \times \mathbf{v}_{\mathbf I}}
                                          {\mathbf{I} \cdot  \mathbf{u}_{\mathbf I}}
                               \otimes \hat{\sigma}.
\end{equation}
It manifests a very interesting property that the component of $\hat{\mathbf j}_\mathbf{I}$ in the direction of $\mathbf I$ is equal to the component of $\hat{\boldsymbol \lambda}$ in the same direction:
$\hat{\mathbf j}_\mathbf{I} \cdot \mathbf{I}=\hat{\boldsymbol \lambda} \cdot \mathbf{I}$.
Note that the first part on the righthand side of Eq. (\ref{TAM-BR}) is not the orbital part of the total angular momentum and the second part is not the spin part, though they are helicity-independent and helicity-dependent, respectively.
With the help of Eqs. (\ref{sI-sigma}), (\ref{CR-lands}), and (\ref{CR-OAM}), it is easy to find
\begin{equation*}
    [(\hat{j}_i)_\mathbf{I}, (\hat{j}_j)_\mathbf{I}]= i \hbar \epsilon_{ijk} (\hat{j}_k)_\mathbf{I},
\end{equation*}
which can be converted into
\begin{equation*}
    [\hat{J}_i, \hat{J}_j]= i \hbar \epsilon_{ijk} \hat{J}_k
\end{equation*}
in the laboratory representation, where
$\hat{\mathbf J}=\hat{\mathbf L}+\hat{\mathbf S}$ is the operator for the total angular momentum in the laboratory representation.

\subsection{Eigen states with spherical harmonics}

Although the total OAM does not satisfy the standard commutation relation (\ref{FCR-L's}), the OAM about the origin of the IRS does. This allows to conveniently give a complete orthonormal set of eigenfunctions that are in the form of spherical harmonics.
Letting be $\lambda$, $\mu$, and $\omega_0$ the canonical quantum numbers that correspond to the eigenvalues of $\hat{\lambda}^2$, $\hat{\lambda}_z$, and $\omega$, respectively, such a complete orthonormal set of eigenfunctions in the laboratory representation can be written as
\begin{equation*}
    \mathbf{f}_{\mathbf{I}, \sigma \lambda \mu \omega_0}
      =\varpi_\mathbf{I} \tilde{\alpha}_\sigma Y_{\lambda \mu} (\mathbf{w})
       \delta(\omega-\omega_0) \exp(-i \omega t),
\end{equation*}
in accordance with Eqs. (\ref{f-EF}) and (\ref{EF}), where
$Y_{\lambda \mu} (\mathbf{w})$
are the spherical harmonics,
\begin{equation*}
    Y_{\lambda \mu} (\mathbf{w})
   =\left\{ \frac{2\lambda +1}{4 \pi} \frac{(\lambda-\mu)!}{(\lambda+\mu)!} \right\}^{1/2}
    P_{\lambda}^{\mu} (\cos \vartheta) e^{i \mu \varphi},
\end{equation*}
which satisfy the following eigenvalue equations,
\begin{eqnarray*}
  \hat{\lambda}^2 Y_{\lambda \mu} &=& \lambda(\lambda+1) \hbar^2 Y_{\lambda \mu},
                                      \quad \lambda=0, 1, 2... \\
  \hat{\lambda}_z Y_{\lambda \mu} &=& \mu \hbar Y_{\lambda \mu},
                                      \quad \mu= 0, \pm 1, \pm 2 ... \pm \lambda,
\end{eqnarray*}
and $\vartheta$ and $\varphi$ are, respectively, the polar and azimuthal angles of $\mathbf w$ in spherical polar coordinates.
It is pointed out that so obtained eigenfunctions are different from the electric or magnetic multipole waves that are also in the form of spherical harmonics \cite{Akhiezer, Cohen, Stratton}.
On one hand, the former describes the eigenstate of the helicity and the latter the eigenstate of the parity. On the other hand, the former is the eigenfunction of the OAM about the barycenter and the latter is the eigenfunction of the total angular momentum.
Roughly speaking, the constant unit vector $\mathbf I$ in the former case is replaced in the latter case with the radial unit vector in the position space as was explicitly shown in Ref. \cite{Stratton}.

\section{Conclusions and remarks}\label{remarks}

In conclusion, the Berry degree of freedom that is expressible in the form of a constant unit vector is identified upon converting the constraint of transversality condition into a quasi unitary matrix. The action of the Berry potential on the helicity of the photon determines the IRS of the photon.
Because the position of the photon in its IRS is canonically conjugate to the momentum, the Berry degree of freedom is endowed with observable quantum effects that show up through the barycenter \cite{Yang} and depend on the helicity \cite{Li09-2, Hosten}. As a result, the position of the photon in the LRS is noncommutative.
In addition, the constraint of transversality condition means that the spin of the photon is aligned with the wavevector direction. Hence only the helicity can be the intrinsic degree of freedom in association with the spin.
But the helicity does not behave intrinsic in the LRS. Only in the IRS does it behave intrinsic.
On the basis of these analyses, the commutation relations of the spin and OAM are derived without having to make use of the second quantization.

It is seen in Section \ref{introducingI} that the Berry degree of freedom is different from the classical gauge degree of freedom. This can also be seen in a different way.
Mathematically, the Berry potential (\ref{AI}) and the Berry degree of freedom arise from the transversality condition (\ref{TC}), which is equivalent to the pair of divergenceless Maxwell's equations:
\begin{equation}\label{pair1}
    \nabla \cdot \mathbf{E}=0, \quad \nabla \cdot \mathbf{H}=0.
\end{equation}
But the electromagnetic potentials and thus the classical gauge degree of freedom arise from the second equation of the above pair and the second one of the following coupled pair:
\begin{equation}\label{pair2}
    \varepsilon_0 \frac{\partial \mathbf{E}}{\partial t}= \nabla \times \mathbf{H}, \quad
    \mu_0 \frac{\partial \mathbf{H}}{\partial t}        =-\nabla \times \mathbf{E}.
\end{equation}
So the Berry degree of freedom originates quite differently than the classical gauge degree of freedom.

Because the Berry potential (\ref{AI}) corresponds to a $\mathbf k$-space ``magnetic monopole'' the singular line of which is along the direction of the Berry degree of freedom, the quantum effects of the Berry degree of freedom mean that the ``location'' of the monopole's singular line is physically observable. This situation is analogous to the vector Aharonov-Bohm effect in real space \cite{Aha, Li1996}.
It seems that the photon in free space or vacuum is not truly ``free'' from the quantum-mechanical point of view. It is ``acted on'' by the field of the $\mathbf k$-space ``magnetic monopole''. The refraction process manifesting the spin Hall effect \cite{Hosten} is a real physical action that changes the ``location'' of the monopole's singular line with respect to the propagation direction \cite{Li09-2}.

Eq. (\ref{CR-xI}) or (\ref{CR-X}) indicates that due to the existence of its helicity, the photon is nonlocal in the LRS. We have shown that this result is implicitly expressed in the form of transversality condition (\ref{TC}).
As is known~\cite{Pryce}, the transversality condition (\ref{TC}), which is equivalent to the pair of divergenceless equations (\ref{pair1}), can be written as a constraint equation about a six-component wavefunction. With its help, the corresponding dynamical equation that is obtained from the Schr\"{o}dinger equation (\ref{Schrodinger-eq-L}) or, equivalently, from the pair of coupled equations (\ref{pair2}) is relativistically covariant.
This means that the nonlocality of the photon with non-vanishing helicity should be regarded as a quantum-mechanical property of the relativistic particle that is expressed by the constraint equation.

\section*{Acknowledgments}

The author is indebted to Vladimir Fedoseyev and Zihua Xin for their helpful discussions. This work was supported in part by the National Natural Science Foundation of China (60877055).

\end{document}